\documentclass[10pt,conference]{IEEEtran}
\IEEEoverridecommandlockouts

\usepackage[]{hyperref}
\usepackage{cite}
\usepackage{amsmath,amssymb,amsfonts}
\usepackage{algorithmic}

\usepackage{graphicx}
\usepackage{textcomp}

\usepackage{algorithm}

\usepackage{kotex}
\usepackage{makecell}
\usepackage{array}
\usepackage[table]{xcolor}
\usepackage{comment}
\usepackage{xspace}

\usepackage{subfig}
\usepackage{float}

\begin{document}

\title{FCDP: Fully Cached Data Parallel for Communication-Avoiding Large-Scale Training}

\author{
\IEEEauthorblockN{Gyeongseo Park, Eungyeong Lee, Song-woo Sok, Myung-Hoon Cha,\\Kwangwon Koh, Baik-Song An, Hongyeon Kim, Ki-Dong Kang$^{*}$}
\thanks{$^{*}$Ki-Dong Kang is the corresponding author (kd\_kang@etri.re.kr).}
\IEEEauthorblockA{Electronics and Telecommunications Research Institute (ETRI)\\
\{gspark, ieg95, swsok, mhcha, kwangwon.koh, bsahn, kimhy, kd\_kang\}@etri.re.kr}
}

\def\arch{\texttt{FCDP}\xspace}
\def\archflow{\texttt{FCDP-Sched}\xspace}
\def\archmem{\texttt{FCDP-Cache}\xspace}
\def\archcomm{\texttt{FCDP-Comm}\xspace}

\def\eg{\textit{e.g.,}}
\def\ie{\textit{i.e.,}}
\newcommand{\niparagraph}[1]{\noindent \textbf{#1}}

\maketitle 
\pagestyle{plain} 

\begin{abstract}
	Training billion-parameter models requires distributing model states across GPUs using fully sharded data parallel (i.e., ZeRO-3).
	While ZeRO-3 succeeds on clusters with high-bandwidth NVLink and InfiniBand interconnects, researchers with commodity hardware face severe inter-node \texttt{all-gather} bottlenecks.
	Existing optimizations take two approaches: GPU-memory caching (MiCS, ZeRO++) trades memory capacity for reduced communication, triggering out-of-memory failures on large models; host memory offloading (ZeRO-Offload, ZeRO-Infinity) extends capacity but degrades throughput due to PCIe overhead.
	We observe that on bandwidth-limited clusters, host memory can serve not as an overflow tier but as a fast caching layer that outperforms inter-node communication.
	Based on this insight, we propose \arch, which eliminates redundant inter-node communication while preserving ZeRO-3's minimal GPU memory footprint.
	\arch caches forward-pass parameters in host memory and reuses them during the backward pass via fast intra-node \texttt{all-gather}, reducing inter-node \texttt{all-gather} by 50\%.
	For parameter-efficient fine-tuning (PEFT), \arch selectively communicates only trainable parameters to maximize caching, reducing inter-node traffic by over 99\%.
	In our commodity cluster setup, \arch achieves up to 100$\times$ higher throughput than ZeRO-3 and 51$\times$ higher than ZeRO++, while maintaining ZeRO-3's maximum batch size.
\end{abstract}

\section{Introduction}
\label{sec:intro}
As language models scale to billions of parameters, their memory footprint exceeds single-GPU capacity, necessitating distributed training strategies.
The Zero Redundancy Optimizer (ZeRO)~\cite{rajbhandari2020zero} has emerged as the de facto approach for such large-scale training. ZeRO-3 shards parameters, gradients, and optimizer states across all GPUs and reconstructs them on demand via \texttt{all-gather} operations.
This full sharding reduces per-GPU memory for model states by approximately a factor of the number of GPUs,
enabling models that far exceed single-device capacity.

However, ZeRO-3's reliance on frequent \texttt{all-gather} operations creates a critical bottleneck on commodity hardware.
While NVIDIA DGX clusters employ 400--800~Gbps InfiniBand, most practitioners train on public clouds with more restricted networks or low-bandwidth clusters~\cite{zhang2022mics, wang2023zero++}.
Analysis of Microsoft's Philly~\cite{jeon2019analysis} and Alibaba's PAI cluster~\cite{weng2022mlaas} traces shows that non-DGX systems cover 90.4\% of the machine specs~\cite{kim2024tccl}, and the networking infrastructure required for DGX costs over 10$\times$ more than PCIe-based alternatives~\cite{an2024fireflyer}.
On such clusters, ZeRO-3 executes \texttt{all-gather} twice per layer per iteration (once for forward, once for backward), and these operations dominate training time.
Our measurements show that training throughput degrades by up to 5.9$\times$ as inter-node network performance decreases (Section~\ref{sec:motivation}).

Recent optimizations address the communication bottleneck by trading GPU memory for reduced communication.
MiCS~\cite{zhang2022mics} limits sharding to smaller sub-groups of GPUs, reducing the scope of \texttt{all-gather} operations.
ZeRO++~\cite{wang2023zero++} caches parameters in GPU memory after the forward pass, eliminating backward-pass \texttt{all-gather}.
Both approaches reduce inter-node traffic but sacrifice ZeRO-3's memory efficiency, triggering out-of-memory failures on large models.
An alternative line of work offloads model states to host memory.
ZeRO-Offload~\cite{ren2021zerooffload} and ZeRO-Infinity~\cite{rajbhandari2021zeroinfinity} treat host memory as an overflow tier for capacity extension, enabling larger models at the cost of additional PCIe transfers that degrade throughput.
These systems prioritize fitting models over accelerating training.
We observe that both approaches overlook a key opportunity: on bandwidth-limited clusters, host memory can serve not as an overflow tier but as a fast caching layer that outperforms inter-node communication.

\begin{table*}[!t]
\centering
\small
\caption{Comparison of distributed training systems. $W$: model parameters, $W_t$: trainable parameters, $G$: total GPUs, $S$: subgroups.}
\setlength{\extrarowheight}{3pt}
\renewcommand{\arraystretch}{1.3}
\resizebox{\textwidth}{!}{%
\begin{tabular}{l|>{\centering\arraybackslash}m{1.2cm}>{\centering\arraybackslash}m{1.2cm}>{\centering\arraybackslash}m{1.2cm}|>{\centering\arraybackslash}m{2cm}|>{\centering\arraybackslash}m{2cm}>{\centering\arraybackslash}m{1.2cm}|m{6.5cm}}
\hline\hline
& \multicolumn{3}{c|}{\textbf{Capabilities}} & \textbf{GPU} &
\multicolumn{2}{c|}{\textbf{Inter-node Comm.}} & \\
\cline{2-4} \cline{6-7}
\textbf{System} &
\textbf{Full}\newline\textbf{Shard} &
\textbf{Cache}\newline\textbf{Tier} &
\textbf{PEFT}\newline\textbf{Aware} &
\textbf{Memory} &
\textbf{Fwd AG}&
\textbf{Bwd AG}&
\textbf{Key Characteristics} \\
\hline

ZeRO-2~\cite{rajbhandari2020zero} &
\cellcolor{red!18}X &
\cellcolor{gray!10}{--} &
\cellcolor{red!18}X &
\cellcolor{red!18}$W$ ($G\times$) &
\cellcolor{green!18}0 &
\cellcolor{red!18}$W$ &
{\footnotesize Params replicated; grad/opt states sharded; additional $W$ sync after optimizer step.} \\
\hline

ZeRO-3~\cite{rajbhandari2020zero} &
\cellcolor{green!18}O &
\cellcolor{gray!10}{--} &
\cellcolor{red!18}X &
\cellcolor{green!18}$W/G$ (1$\times$) &
\cellcolor{red!18}$W$ &
\cellcolor{red!18}$W$ &
{\footnotesize Full sharding baseline; AG required before each Fwd/Bwd computation.} \\
\hline

MiCS~\cite{zhang2022mics} &
\cellcolor{yellow!25}$S$ &
\cellcolor{gray!10}{--} &
\cellcolor{red!18}X &
\cellcolor{yellow!25}$WS/G$ ($S\times$) &
\cellcolor{red!18}$W$ &
\cellcolor{red!18}$W$ &
{\footnotesize Shards within subgroups; trades memory for reduced communication scope.} \\
\hline

ZeRO++~\cite{wang2023zero++} &
\cellcolor{green!18}O &
\cellcolor{yellow!25}GPU &
\cellcolor{red!18}X &
\cellcolor{yellow!25}$WS/G$ ($S\times$) &
\cellcolor{yellow!25}$W$\newline{\scriptsize(+qwZ: $\frac{W}{2}$)} &
\cellcolor{green!18}0 &
{\footnotesize GPU cache eliminates Bwd-AG; quantized Fwd-AG; requires extra GPU memory for cache.} \\
\hline

\arch &
\cellcolor{green!18}O &
\cellcolor{green!18}CPU &
\cellcolor{green!18}O &
\cellcolor{green!18}$W/G$ (1$\times$) &
\cellcolor{green!18}$W_t\approx 0$ (LoRA) &
\cellcolor{green!18}0 &
{\footnotesize Host cache (no extra GPU mem); PEFT-aware comm skips frozen params; maintains full sharding.} \\
\hline\hline
\end{tabular}%
}
\label{tab:intro_comparison}
\end{table*}

\noindent\textbf{(C1) Host memory can outperform inter-node communication on bandwidth-limited clusters.}
On clusters where inter-node bandwidth is constrained, transferring parameters from host memory to GPU via PCIe can be faster than fetching them from remote GPUs via the network.
Our measurements confirm that local CPU-to-GPU transfers significantly outperform inter-node transfers on commodity hardware (Section~\ref{sec:motivation}).
This reveals an underutilized opportunity: parameters reconstructed during the forward pass can be cached in host memory and reloaded faster than re-fetching them via inter-node \texttt{all-gather}.

Realizing this potential, however, requires addressing three challenges.
First, host memory introduces CPU-to-GPU transfer overhead that must be carefully scheduled to avoid stalling the training pipeline.
Second, a static caching policy is insufficient: ZeRO-3 minimizes GPU memory but requires inter-node communication for every layer, while ZeRO++ caches on GPU but consumes additional memory that reduces the maximum batch size.
Third, parameter updates after each optimizer step invalidate the cache and require re-gathering across all nodes, raising the question of whether this overhead can be fundamentally reduced.

To address these challenges, we propose \arch, a distributed training system that uses host memory as a first-class caching tier.
Unlike ZeRO++, which caches in GPU memory, \arch preserves ZeRO-3's minimal GPU footprint while achieving comparable communication reduction.
The following contributions address each challenge in turn.

\noindent\textbf{(C2) Forward-pass parameters can be reused in the backward pass, eliminating redundant communication.}
ZeRO-3 performs inter-node \texttt{all-gather} twice per layer (once for forward, once for backward), even though the same parameters are needed both times.
We exploit this redundancy by caching parameters to host memory after the forward-pass \texttt{all-gather}; during the backward pass, each GPU loads its cached shard locally and reconstructs full parameters via fast intra-node \texttt{all-gather}.
This scheduling eliminates backward-pass inter-node communication, reducing inter-node all-gather traffic by 50\%.

\noindent\textbf{(C3) Dynamic memory pressure enables adaptive caching without sacrificing capacity.}
Static caching policies force a trade-off: GPU caching (ZeRO++) reduces communication but consumes memory, while host-only caching preserves memory but incurs PCIe overhead on every access.
We observe that GPU memory pressure fluctuates during training, creating opportunities for selective on-device caching.
\arch monitors available GPU memory and adaptively places parameters: on-device when headroom exists, in host memory under pressure.
This guarantees worst-case memory usage identical to ZeRO-3 while opportunistically eliminating PCIe transfers.

\noindent\textbf{(C4) Parameter classification enables communication elimination for frozen weights.}
Standard systems treat all parameters identically, re-gathering them every iteration regardless of whether they change.
We observe that parameter-efficient fine-tuning (PEFT) workloads such as LoRA~\cite{hu2022lora} partition parameters into frozen base weights and trainable adapters, while frozen parameters remain constant throughout training.
By classifying parameters at initialization, \arch gathers frozen weights once and caches them indefinitely, restricting inter-node \texttt{all-gather} to trainable adapters only.
For LoRA fine-tuning where 99\% of parameters are frozen, this insight reduces inter-node traffic by over 99\%.

\noindent\textbf{(C5) We achieve higher throughput than ZeRO++ while maintaining ZeRO-3's memory efficiency.}
We evaluate \arch on a 4-node cluster under various commodity network configurations, representative of hardware available to most researchers.
For full fine-tuning of GPT-style models (10B--30B parameters), \arch achieves up to 41.3\% higher throughput than ZeRO-3 and 2$\times$ higher than ZeRO++ while maintaining identical maximum batch size.
For LoRA fine-tuning, \arch achieves up to 100$\times$ higher throughput than ZeRO-3 and 51$\times$ higher than ZeRO++ by eliminating inter-node communication for frozen parameters.

Table~\ref{tab:intro_comparison} summarizes the trade-offs across distributed training systems.
ZeRO-3 achieves minimal GPU memory through full sharding, but requires inter-node \texttt{all-gather} for both forward and backward passes.
MiCS and ZeRO++ reduce communication by caching parameters, but this caching consumes additional GPU memory, limiting the maximum trainable batch size and causing out-of-memory failures on large models.
\arch breaks this trade-off by caching in host memory instead of GPU memory: it eliminates backward-pass inter-node \texttt{all-gather} like ZeRO++ while preserving ZeRO-3's minimal GPU footprint.
Furthermore, \arch is PEFT-aware. For LoRA workloads, it caches frozen weights indefinitely and applies inter-node communication only to trainable adapters (typically $<$1\% of parameters).

\section{Background}
\label{sec:background}

\begin{figure*}[!t]
\centering
\includegraphics[width=54em]{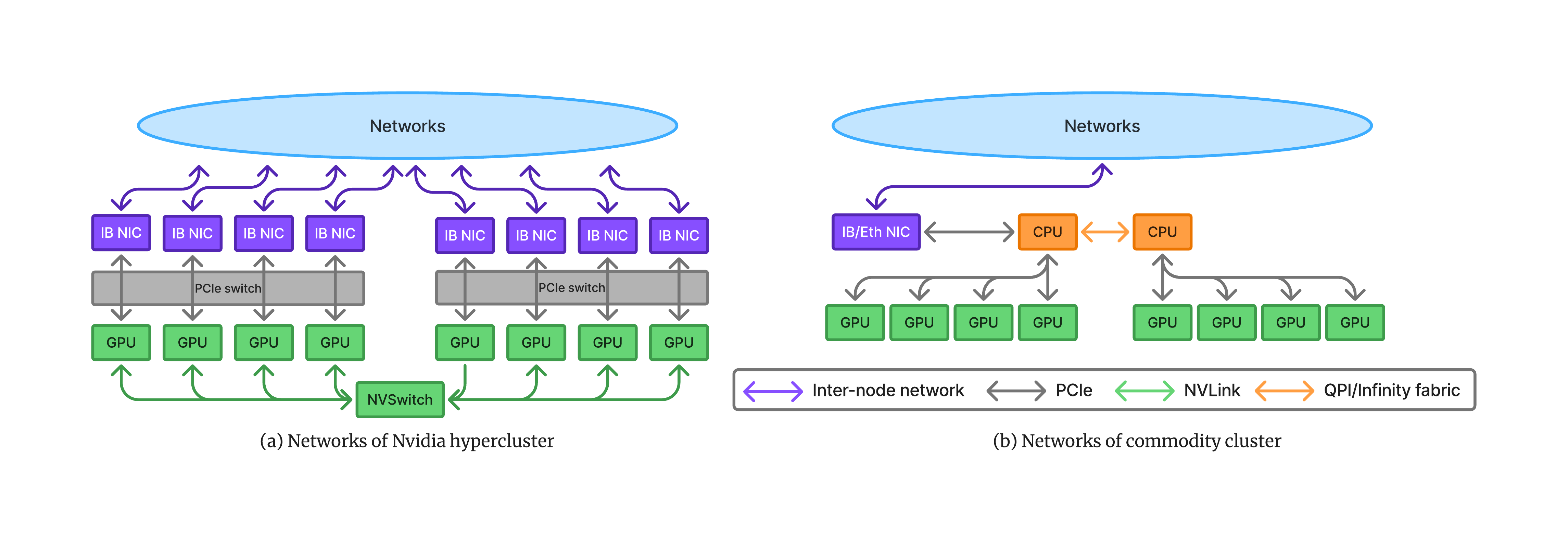}
\vspace{-3.0em}
\caption{Comparison of hypercluster and commodity cluster.}
\label{fig:background}
\end{figure*}

\subsection{ZeRO-Style Data Parallelism and Communication}

ZeRO-style data parallelism~\cite{rajbhandari2020zero,wang2023zero++} shards model states across GPUs while preserving the single-GPU programming abstraction.
Model states comprise parameters, gradients, and optimizer states.
ZeRO defines three sharding stages: ZeRO-1 shards only optimizer states; ZeRO-2 additionally shards gradients; and ZeRO-3 shards all three components across all GPUs, achieving the minimum per-GPU memory footprint and making it the default choice for training the largest models.

The trade-off of full sharding is communication.
Since ZeRO-3 does not replicate parameters, it must reconstruct the full parameter tensor via \texttt{all-gather} before every forward and backward computation, and synchronize gradients via \texttt{reduce-scatter}, resulting in three collective operations per layer per iteration.
On bandwidth-limited commodity clusters, this repeated parameter reconstruction becomes the dominant runtime cost.
A detailed theoretical analysis of per-system memory and communication trade-offs is provided in Section~\ref{sec:discussion}.

\subsection{Cluster Architecture and Communication Cost}

The cost of ZeRO-3's collectives depends heavily on cluster topology.
Figure~\ref{fig:background} contrasts two representative configurations.
The hypercluster (DGX/HGX-style) connects GPUs via NVSwitch within a node and fans out to multiple InfiniBand NICs for inter-node traffic; both intra-node and inter-node paths are heavily provisioned, enabling frequent \texttt{all-gather} without bottlenecks.
The commodity cluster~\cite{an2024fireflyer}, in contrast, places GPUs behind CPU sockets over PCIe with a single NIC per node, so inter-node traffic must traverse the CPU interconnect, PCIe root complex, and the lone NIC. This narrow path limits effective bandwidth.
Table~\ref{tab:bandwidth} quantifies the gap: intra-node bandwidth (NVLink or PCIe) is 2.5--16$\times$ higher than inter-node bandwidth.
This bandwidth gap makes ZeRO-3's inter-node \texttt{all-gather} the dominant cost: as node count $N$ increases, GPUs spend more time waiting for parameters and less time computing, making inter-node communication the primary scaling bottleneck.

\begin{table}[b!]
\centering
\small
\caption{Theoretical bandwidth in commodity clusters.}
\label{tab:bandwidth}
\begin{tabular}{lcc}
\hline
\textbf{Path} & \textbf{Bandwidth} & \textbf{Relative} \\
\hline
PCIe 4.0 $\times$16 (GPU--CPU) & 32 GB/s & 2.5$\times$ \\
NVLink 3.0 $\times$4 (GPU--GPU) & 200 GB/s & 16$\times$ \\
100 Gbps Ethernet (inter-node) & 12.5 GB/s & 1$\times$ \\
\hline
\end{tabular}
\end{table}

\subsection{Parameter-Efficient Fine-Tuning with LoRA}

As models grow, full fine-tuning becomes prohibitively expensive, and PEFT addresses this by training only a small subset of parameters while freezing the rest.
Low-Rank Adaptation (LoRA)~\cite{hu2022lora} is a prominent PEFT method that adds trainable low-rank matrices to selected layers: for a frozen weight matrix $\mathbf{W}_0 \in \mathbb{R}^{d \times k}$, LoRA adds $\Delta\mathbf{W} = BA$ where $B \in \mathbb{R}^{d \times r}$, $A \in \mathbb{R}^{r \times k}$, and rank $r \ll \min(d, k)$.
Only $A$ and $B$ are trained; $\mathbf{W}_0$ remains frozen throughout training, dramatically reducing trainable parameters. For example, rank-16 LoRA on LLaMA-7B trains only 13M parameters, just 0.2\% of the 6.7B base parameters.
Let $W_t$ denote trainable parameters and $W_f$ denote frozen parameters; in LoRA, $W_t \ll W_f$.
Under ZeRO-3, however, all parameters are treated uniformly: every iteration, the full model is reconstructed via \texttt{all-gather} regardless of whether parameters are trainable or frozen.
Communication volume therefore scales with total model size $W = W_t + W_f$, not trainable size $W_t$.

\section{Motivation}
\label{sec:motivation}

%
%
%
%
%
%
%
%

\noindent
The previous section established that ZeRO-3's parameter reconstruction dominates communication cost, that commodity clusters suffer from limited inter-node bandwidth, and that PEFT workloads communicate far more data than necessary.
This section quantifies these inefficiencies and reveals an underutilized opportunity: host memory as a fast caching layer.

\subsection{Communication Dominates Training Time}

\begin{figure}[t]
\centering
\includegraphics[width=1\linewidth]{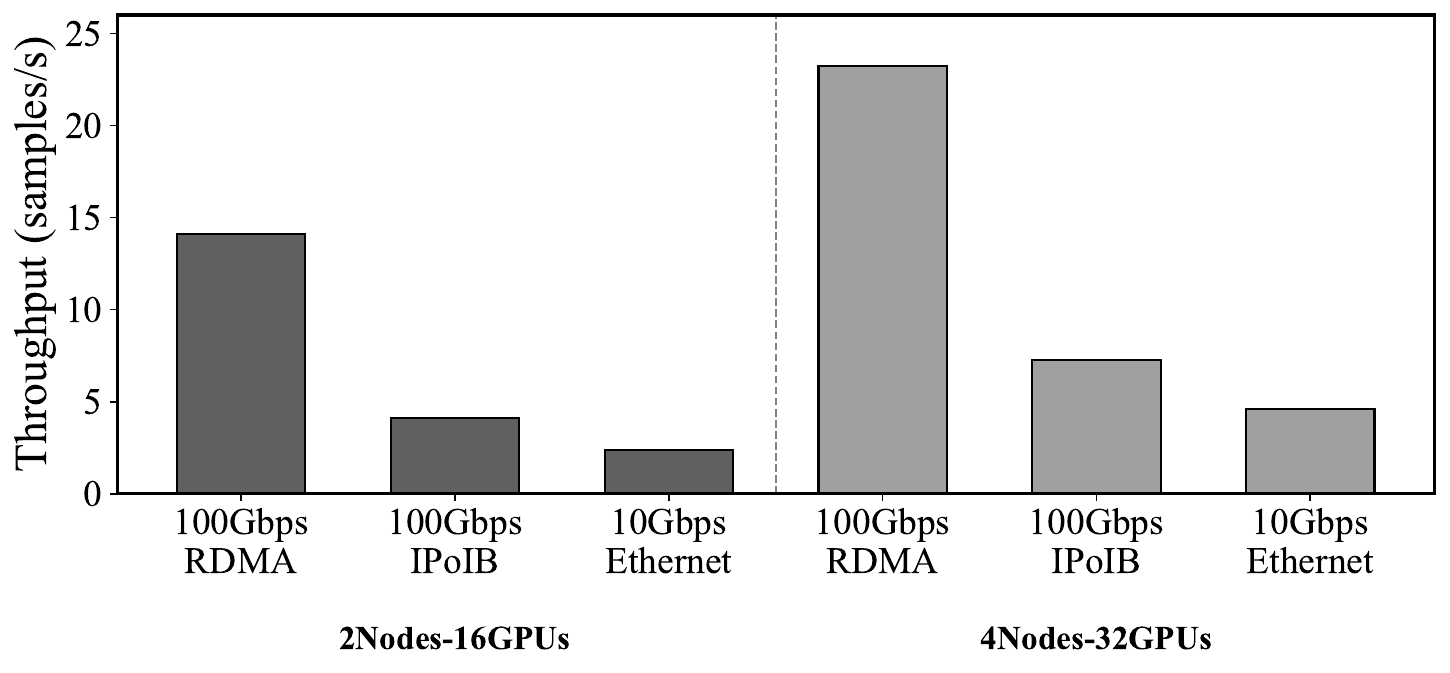}
\caption{ZeRO-3 training throughput (GPT-10B, batch 8) across network configurations.}
\label{fig:motivation_network}
\end{figure}

Commodity clusters offer cost-effective infrastructure, but their inter-node network performance directly constrains training throughput.
To quantify this impact, we measure ZeRO-3 throughput for GPT-10B across three network configurations in Figure~\ref{fig:motivation_network}; setup details in Section~\ref{sec:eval_setup}.
On a 2-node cluster (16 GPUs), throughput reaches 14.1 samples/s with 100Gbps Remote Direct Memory Access (RDMA) but degrades to 4.1 samples/s over IP over InfiniBand (IPoIB) ($3.4\times$ slowdown) and just 2.4 samples/s over 10Gbps Ethernet ($5.9\times$ slowdown).
The degradation scales with cluster size. At 4 nodes (32 GPUs), RDMA achieves 23.2 samples/s while IPoIB and 10Gbps Ethernet yield only 7.3 and 4.6 samples/s ($3.2\times$ and $5.0\times$ slowdowns, respectively).

This degradation stems from ZeRO-3's inter-node \texttt{all-gather} operations: each layer requires two parameter reconstructions per iteration (forward and backward), plus one gradient \texttt{reduce-scatter}.
Unlike hypercluster environments where NVLink and NVSwitch provide 900GB/s+ interconnects that sustain these collectives efficiently, commodity clusters expose \texttt{all-gather} as the dominant bottleneck.
Each layer's parameters must traverse the inter-node network twice per iteration, forcing GPUs to remain idle while waiting for data transfers.

\subsection{Memory-Communication Trade-off Limits Scalability}

Existing systems address communication overhead by caching parameters in GPU memory, but this approach imposes a memory penalty that limits scalability.
On A40-48GB GPUs running GPT-15B (batch size 8, bf16), ZeRO-3 consumes 34.06\,GB per GPU, while ZeRO++ requires 40.92\,GB due to its node-level parameter cache, a 6.86\,GB increase (20.1\%) that directly reduces the memory available for activations and batch size.

This creates a dilemma: scaling to larger models or batch sizes under ZeRO++ risks out-of-memory failures, yet reverting to ZeRO-3 reintroduces the full communication overhead.
MiCS faces a similar constraint. Confining sharding to GPU subgroups reduces communication scope but increases per-GPU memory proportionally to the subgroup size.
In commodity environments where both inter-node bandwidth and GPU memory are limited, neither approach resolves both bottlenecks simultaneously.
This motivates an alternative: leveraging host memory as a caching tier to eliminate redundant inter-node communication without consuming GPU memory.

\subsection{Host Memory as Underutilized Resource}

\begin{table}[t]
\centering
\caption{Communication latency for 16GB data transfer across 2 nodes (8 GPUs per node). CPU-to-GPU transfer via PCIe is faster than all inter-node alternatives.}
\label{tab:motivation_comm}
\begin{tabular}{lcc}
\hline\hline
\textbf{Communication Path} & \textbf{Latency (s)} & \textbf{vs.\ PCIe} \\
\hline
\textbf{CPU to GPU (PCIe 4.0)} & \textbf{0.613} & \textbf{1.0$\times$} \\
\hline
InfiniBand 100Gbps RDMA & 0.949 & 1.5$\times$ \\
InfiniBand 100Gbps IPoIB & 3.963 & 6.5$\times$ \\
Ethernet 10Gbps & 6.745 & 11$\times$ \\
Ethernet 1Gbps & 67.66 & 110$\times$ \\
\hline\hline
\end{tabular}
\end{table}

To quantify the opportunity for host-memory caching, we measure the time to transfer 16\,GB of data under different communication paths in Table~\ref{tab:motivation_comm}.
Inter-node \texttt{all-gather} via 100Gbps RDMA takes 0.949\,s, while IPoIB and 10Gbps Ethernet require 3.963\,s and 6.745\,s, respectively.
In contrast, a local CPU-to-GPU transfer via PCIe completes in only 0.613\,s, faster than even InfiniBand RDMA.

This gap reveals an underutilized opportunity.
Current ZeRO-3 workflows perform redundant inter-node \texttt{all-gather} operations every iteration, yet the same parameters could be cached in host memory and reloaded via PCIe at lower latency.
Prior offloading systems treat host memory merely as overflow storage for capacity extension.
We instead identify host memory as a first-class caching layer that actively reduces inter-node traffic.
By caching forward-pass parameters in host memory and reusing them during the backward pass, \arch shifts communication from the slow inter-node path to the faster PCIe path, reducing bandwidth pressure while preserving ZeRO-3's memory efficiency.

\section{Architecture}
\label{sec:architecture}
%
%

\begin{figure*}[t!]
    \centering
    \includegraphics[width=\textwidth]{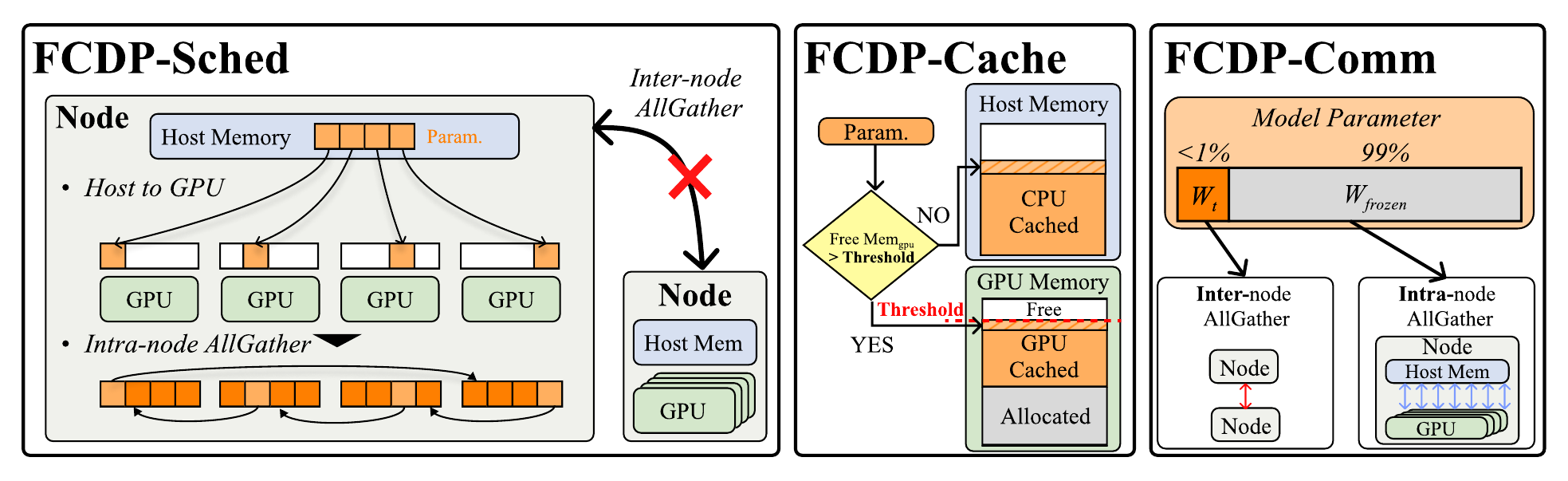}
    \caption{Overview of \arch's three components: \archflow (parameter scheduling), \archmem (adaptive memory placement), and \archcomm (PEFT-aware communication).}
    \label{fig:arch_overview}
\end{figure*}

\subsection{Challenges}
Host memory offers a compelling opportunity for distributed training by providing a local cache that can eliminate slow inter-node communication.
However, realizing this potential requires addressing three challenges that prior systems have not solved together.

\niparagraph{Scheduling complexity.}
Host memory provides large capacity but limited PCIe bandwidth.
Although local PCIe transfers can complete faster than inter-node transfers for large payloads, naively inserting CPU--GPU copies can serialize with collective communication and stall the training pipeline.
The system must carefully schedule when to store parameters to host memory and when to prefetch them back, coordinating these transfers with all-gather operations across both forward and backward passes.

\niparagraph{Static caching trade-offs.}
GPU memory that remains unused after allocating shards, optimizer states, and gradients represents an optimization opportunity.
ZeRO-3 stores only sharded parameters, minimizing footprint but requiring inter-node communication for every layer.
ZeRO++ caches full parameters in GPU memory, reducing communication but consuming substantial memory.
No existing solution dynamically adapts GPU and host memory allocation for parameter caching based on runtime conditions.
A static all-or-nothing approach either wastes available GPU memory or forces unnecessary PCIe transfers when memory pressure is low.

\niparagraph{Update-driven communication.}
When parameters are cached in host memory, optimizer updates invalidate the cache.
Updated parameters must be re-gathered across all GPUs, potentially reintroducing inter-node communication overhead.
For PEFT workloads where the vast majority of parameters remain frozen, this uniform treatment is wasteful: frozen weights never change, yet standard systems re-gather them every iteration as if they were updated.
Without distinguishing frozen from trainable parameters, the communication benefits of host-memory caching cannot be fully realized.

\subsection{Design Overview}
\label{subsec:overview}

We propose \arch (Fully Cached Data Parallel), a distributed training system that uses host memory as a first-class caching tier to reduce inter-node communication while preserving ZeRO-3's memory efficiency.
The key idea is to cache parameters constructed during the forward pass in host memory and reload them via local PCIe transfers before the backward pass, replacing slow inter-node \texttt{all-gather} with faster intra-node data movement.
This approach maintains ZeRO-3's minimal GPU memory footprint while achieving communication reduction comparable to ZeRO++, which requires additional GPU memory for caching.

Figure~\ref{fig:arch_overview} illustrates the three components of \arch that address the challenges.
\textbf{\archflow} addresses \emph{scheduling complexity} by coordinating CPU--GPU transfers with collective operations: it stores parameters to host memory after forward \texttt{all-gather} and prefetches them before backward computation, avoiding pipeline stalls while eliminating inter-node backward \texttt{all-gather}.
\textbf{\archmem} overcomes \emph{static caching trade-offs} by dynamically adapting memory placement at runtime: when GPU headroom exists, parameters remain on-device; under memory pressure, they offload to host memory, avoiding the all-or-nothing limitation of prior approaches.
\textbf{\archcomm} eliminates \emph{update-driven communication} by recognizing that frozen weights never change: it gathers frozen base weights ($W_f$, 99\%) once and caches them indefinitely, applying inter-node \texttt{all-gather} only to trainable adapters ($W_t$, $<$1\%).

\begin{figure*}[t!]
    \centering
    \includegraphics[width=1.0\textwidth]{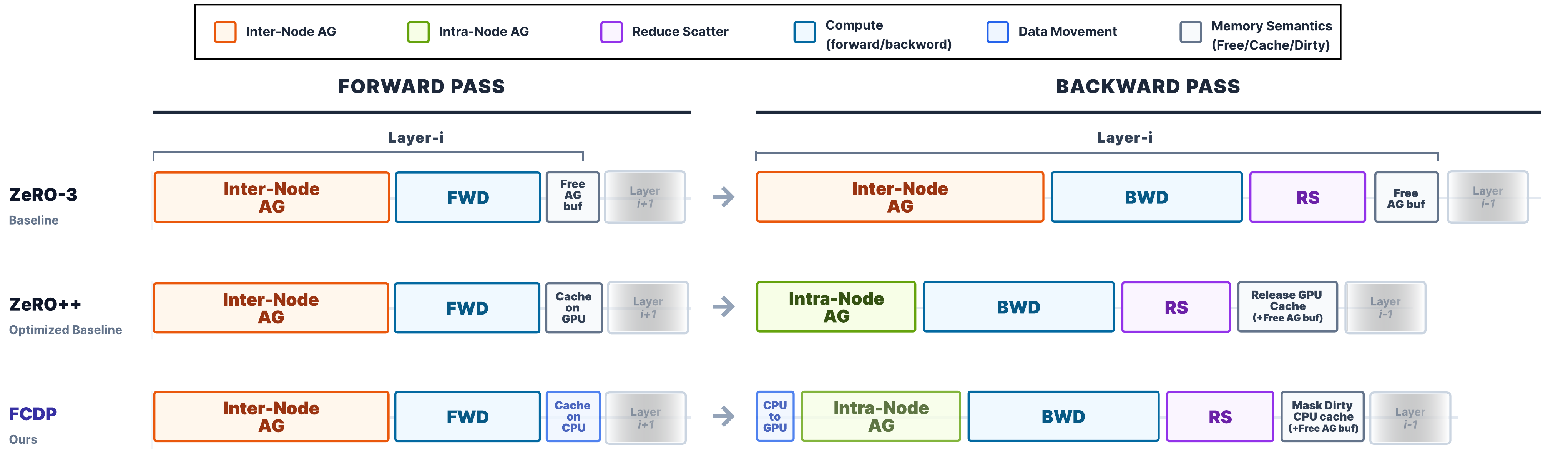}
    \caption{Per-layer execution schedule comparison. ZeRO-3 performs inter-node AG twice; ZeRO++ caches on GPU for intra-node backward AG; \arch caches on CPU, achieving intra-node backward AG without GPU memory overhead.}
    \label{fig:arch_schedule}
\end{figure*}

\subsection{\archflow: Parameter Scheduling}
\label{subsec:archflow}
\archflow eliminates backward-pass inter-node \texttt{all-gather} by caching forward-reconstructed parameters in host memory.
This reduces inter-node communication by 50\% compared to ZeRO-3, while preserving minimal GPU memory footprint. Unlike ZeRO++, which caches on GPU and incurs memory overhead, \archflow caches in host memory.

Figure~\ref{fig:arch_schedule} illustrates the per-layer execution schedule.
During the \textbf{forward pass}, inter-node \texttt{all-gather} reconstructs full parameters from shards, the layer computes forward activations, and \archflow asynchronously copies the parameters to host memory (``Cache on CPU'' in Figure~\ref{fig:arch_schedule}).
During the \textbf{backward pass}, each GPU first loads its $1/N$ shard of cached parameters from host memory via PCIe (``CPU to GPU''), where $N$ is the number of GPUs per node.
The GPUs then perform intra-node \texttt{all-gather} using high-bandwidth NVLink or PCIe interconnects (e.g., ring communication) to exchange shards and reconstruct the full parameters needed for the current layer, all without inter-node traffic.
After gradient computation and \texttt{reduce-scatter}, the CPU cache is marked dirty (``Mask Dirty'') rather than deallocated.
Since optimizer updates modify parameters, the cache becomes outdated; marking it dirty triggers a lazy refresh on the next forward \texttt{all-gather}, which overwrites the cache with fresh values.
This approach avoids host-memory deallocation overhead while guaranteeing correctness (Section~\ref{subsec:archmem}).

\archflow eliminates all backward-pass inter-node \texttt{all-gather} operations, reducing inter-node \texttt{all-gather} traffic by 50\%.
The replacement PCIe transfers overlap with computation, minimizing impact on the critical path.
Unlike ZeRO-3 which performs inter-node \texttt{all-gather} twice per layer, or ZeRO++ which caches on GPU and adds memory overhead, \archflow caches in host memory. This achieves the same communication reduction as ZeRO++ while preserving ZeRO-3's minimal GPU footprint.

\subsection{\archmem: Memory Management}
\label{subsec:archmem}
\archmem manages parameter caching across GPU and host memory tiers, adapting placement based on runtime memory pressure while optimizing host-memory transfers through NUMA-aware pinned buffers.

\noindent
\textbf{Adaptive GPU Caching.}
When GPU memory headroom exists, retaining parameters on-device eliminates PCIe transfer overhead entirely.
\archmem makes this decision at the ``Cache on CPU'' point in Figure~\ref{fig:arch_schedule}: after forward computation completes for each layer, it checks GPU memory utilization against a threshold $\tau$.
If utilization falls below $\tau$, the layer's parameters remain on GPU for direct reuse in the backward pass, skipping both the GPU$\to$CPU store and CPU$\to$GPU load.
If utilization exceeds $\tau$, parameters offload to host memory following the baseline \archflow schedule.
This per-layer adaptive policy reduces total data movement during small-batch workloads while ensuring worst-case behavior identical to ZeRO-3 under memory pressure.
Setting $\tau$ near 100\% causes aggressive GPU caching that minimizes PCIe transfers but risks out-of-memory under large batches; setting $\tau$ near 0\% forces all parameters to host memory, guaranteeing memory safety at the cost of additional PCIe overhead.
The threshold $\tau$ is user-configurable to balance memory headroom with transfer reduction.

\noindent
\textbf{Host Memory Optimization.}
For host-memory storage, \archmem pre-allocates pinned (page-locked) buffers at model initialization and retains them throughout training. Buffers are never deallocated between iterations.
This pool-based approach eliminates per-allocation latency and enables parameter updates in-place without reallocation overhead.
On multi-socket systems, naive allocation can route GPU traffic across QPI/UPI interconnects, degrading effective PCIe bandwidth.
\archmem addresses this by allocating pinned buffers from the Non-Uniform Memory Access (NUMA) node local to each GPU and pinning training processes to CPU cores in the same NUMA domain.
This NUMA-aware placement maximizes PCIe bandwidth for CPU--GPU transfers.

\subsection{\archcomm: PEFT-Aware Communication}
\label{subsec:archcomm}
While \archflow eliminates backward-pass inter-node communication, forward passes still require inter-node \texttt{all-gather} every iteration.
\archcomm addresses this for PEFT workloads, reducing per-iteration inter-node communication by orders of magnitude.
The key insight is that frozen parameters (99\%+ of model weights in typical LoRA configurations) never change after initialization, so they need only be gathered once and cached permanently rather than re-gathered every iteration like trainable adapters.

Algorithm~\ref{alg:peft-aware} illustrates the training loop.
Each parameter $p$ has two sharded representations.
$p^{\text{inter}}$ is the ZeRO-3 shard that each GPU holds under global partitioning, and $p^{\text{intra}}$ is the intra-node shard loaded from the host cache $H$ for intra-node reconstruction.
The flag $\mathit{dirty}(p)$ tracks whether $H[p]$ is stale.
\textsc{AllGather}\textsuperscript{inter} gathers shards across all nodes, while \textsc{AllGather}\textsuperscript{intra} gathers shards within a single node.

All parameters start with $\mathit{dirty}(p)$ set to true (lines 3--4), since no cached values exist in $H$ initially.
The forward-pass logic handles the first gather uniformly across all nodes.
From the second iteration onward, frozen parameters remain clean indefinitely while only trainable parameters are re-marked dirty after optimizer step.

In each forward pass (lines 8--19), layers are processed sequentially.
For every parameter $p$ in layer $l$, the algorithm checks $\mathit{dirty}(p)$ (line 10).
If dirty, an inter-node \texttt{all-gather} over $p^{\text{inter}}$ fetches the latest value and refreshes $H$ (lines 11--12).
If clean, the shard $p^{\text{intra}}$ is loaded from $H$ to the GPU via PCIe (line 14) and reconstructed through intra-node \texttt{all-gather} (line 15).
After the first iteration, only trainable parameters ($<$1\% of weights in typical LoRA configurations) remain dirty, so the vast majority of parameters take the cache-hit path.

The backward pass (lines 22--28) requires no inter-node communication.
All intra-node shards $p^{\text{intra}}$ are loaded from $H$ and reconstructed via intra-node \texttt{all-gather}, followed by a single \texttt{reduce-scatter} for gradient synchronization across nodes (line 29).
After the optimizer step (line 32), only trainable parameters are marked dirty (lines 33--34), triggering a cache refresh in the next forward pass.
Frozen parameters remain clean throughout training, eliminating inter-node communication for 99\%+ of parameters.

\begin{algorithm}[t]
  \caption{PEFT-aware training loop}
  \label{alg:peft-aware}
  \small
  \begin{algorithmic}[1]
    \REQUIRE $L$ layers; $W_f$: frozen params; $W_t$: trainable params
    \ENSURE $p$: full parameter; $\nabla W_t$: trainable gradients
    \ENSURE $p^{\text{inter}}$: ZeRO-3 shard of $p$ (across all GPUs)
    \ENSURE $p^{\text{intra}}$: intra-node shard of $p$ from $H$ (within one node)
    \ENSURE $H$: host cache; $\mathit{dirty}(p)$: true when $H[p]$ is stale
    \STATE
    \STATE \textbf{\#Initialize}
    \FORALL{$p \in W_f \cup W_t$}
      \STATE $\mathit{dirty}(p) \gets$ true
    \ENDFOR
    \STATE
    \STATE \textbf{\#Forward}
    \FOR{$l = 1$ \TO $L$}
      \FORALL{$p \in$ layer $l$}
        \IF{$\mathit{dirty}(p)$}
          \STATE $p \gets$ \textsc{AllGather}\textsuperscript{inter}($p^{\text{inter}}$)
          \STATE $H[p] \gets p$; \; $\mathit{dirty}(p) \gets$ false \hfill \text{\#gpu to host}
        \ELSE
          \STATE $p^{\text{intra}} \gets H[p]$  \hfill \text{\#host to gpu}         
          \STATE $p \gets$ \textsc{AllGather}\textsuperscript{intra}($p^{\text{intra}}$)
        \ENDIF
      \ENDFOR
      \STATE \textsc{ComputeForward}($l$)
    \ENDFOR
    \STATE
    \STATE \textbf{\#Backward}
    \FOR{$l = L$ \TO $1$}
      \FORALL{$p \in$ layer $l$}
        \STATE $p^{\text{intra}} \gets H[p]$
        \STATE $p \gets$ \textsc{AllGather}\textsuperscript{intra}($p^{\text{intra}}$)
      \ENDFOR
      \STATE \textsc{ComputeBackward}($l$)
    \ENDFOR
    \STATE \textsc{ReduceScatter}\textsuperscript{inter}($\nabla W_t$)
    \STATE
    \STATE \textbf{\#Update}
    \STATE \textsc{OptimizerStep}($W_t$)
    \FORALL{$p \in W_t$}
      \STATE $\mathit{dirty}(p) \gets$ true
    \ENDFOR
  \end{algorithmic}
\end{algorithm}
\section{Evaluation}

%
%
%
%
%
%

We evaluate \arch on multi-node commodity clusters and compare it against ZeRO-3~\cite{rajbhandari2020zero} and ZeRO++~\cite{wang2023zero++}.
Our evaluation focuses on two key questions:
(1)~Does \arch reduce inter-node communication overhead without increasing memory footprint, and (2)~how does \arch perform compared to ZeRO++ under memory-constrained configurations?
We measure training throughput, evaluate memory efficiency, and characterize communication behavior across a range of model sizes and cluster configurations.

\subsection{Methodology}
\label{sec:eval_setup}

\noindent\textbf{Hardware Configuration.}
Our testbed consists of four nodes, each equipped with dual AMD EPYC 7313 16-core processors, 512\,GB of host memory, and eight NVIDIA A40 GPUs.
Each A40 GPU provides 48\,GB of on-device memory.
Within a node, GPUs are connected in pairs via NVLink bridges, and all GPUs attach to two CPU sockets over PCIe Gen4 x16 links.
The four nodes are interconnected via InfiniBand 100\,Gbps EDR, forming a 32-GPU cluster.
This configuration represents a commodity-style topology where inter-node bandwidth is substantially lower than the aggregate intra-node GPU--GPU bandwidth provided by NVLink.

\noindent\textbf{Software Stack.}
We implement \arch as an extension to DeepSpeed version 0.16.2 (approximately 3,200 lines of Python) using PyTorch 2.3.0 and Python 3.10.12.
Host-memory buffers use CUDA pinned memory for DMA transfers at full PCIe bandwidth, allocated from the NUMA node local to each GPU on multi-socket systems.
GPU-to-CPU copies run on a dedicated CUDA stream to overlap with forward computation.
All experiments run in full fine-tuning mode with mixed-precision training (FP16 activations and gradients, FP32 master weights).
\arch requires approximately $2W$ bytes of host memory per node for caching (20\,GB for a 10B model in FP16).

\noindent\textbf{Models and Baselines.}
We evaluate on GPT-2-XL--based models from Hugging Face, scaled to 10\,B, 15\,B, 20\,B, 25\,B, and 30\,B parameters by increasing the number of layers while preserving the per-layer structure of GPT-2-XL.
Table~\ref{table:model_configs} lists the detailed configurations.
All experiments use the SQuAD dataset~\cite{rajpurkar2016squad}.
We compare \arch against two baselines: ZeRO-3, which shards all model states across GPUs and reconstructs parameters via inter-node \texttt{all-gather}, and ZeRO++, which caches a node-local replica of model parameters in GPU memory to eliminate backward-pass inter-node communication.

\begin{table}[t]
\centering
\caption{Model configurations derived from GPT-2-XL.}
\label{table:model_configs}
\begin{tabular}{lrrrrr}
\hline
\textbf{Model} & \textbf{Params} & \textbf{Layers} & \textbf{Hidden} & \textbf{Heads} & \textbf{Vocab} \\
\hline
GPT-10B  & 10B & 40  & 4800 & 40 & 50257 \\
GPT-15B  & 15B & 40  & 5760 & 45 & 50257 \\
GPT-20B  & 20B & 40  & 6656 & 52 & 50257 \\
GPT-25B  & 25B & 39  & 7168 & 56 & 50257 \\
GPT-30B  & 30B & 40  & 7936 & 62 & 50257 \\
\hline
\end{tabular}
\end{table}

\noindent\textbf{Metrics.}
We report training throughput in samples per second.
We also measure inter-node communication volumes using profiling hooks in DeepSpeed.

\begin{figure*}[t]
\centering
\subfloat[2nodes-16GPUs]
{
\includegraphics[
width=25em,
trim=6 0 0 0, 
clip
]{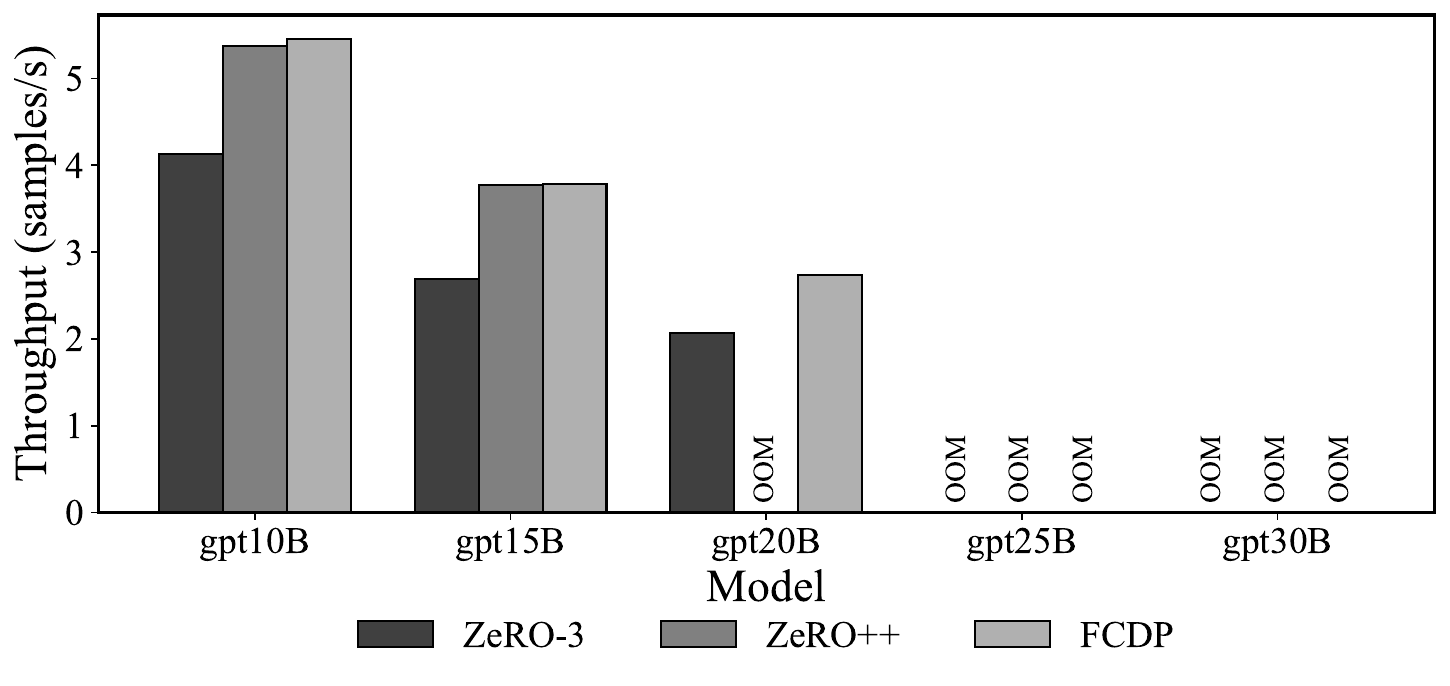}
}
\subfloat[4nodes-32GPUs]
{
\includegraphics[
width=25em,
trim=6 0 0 0, 
clip
]{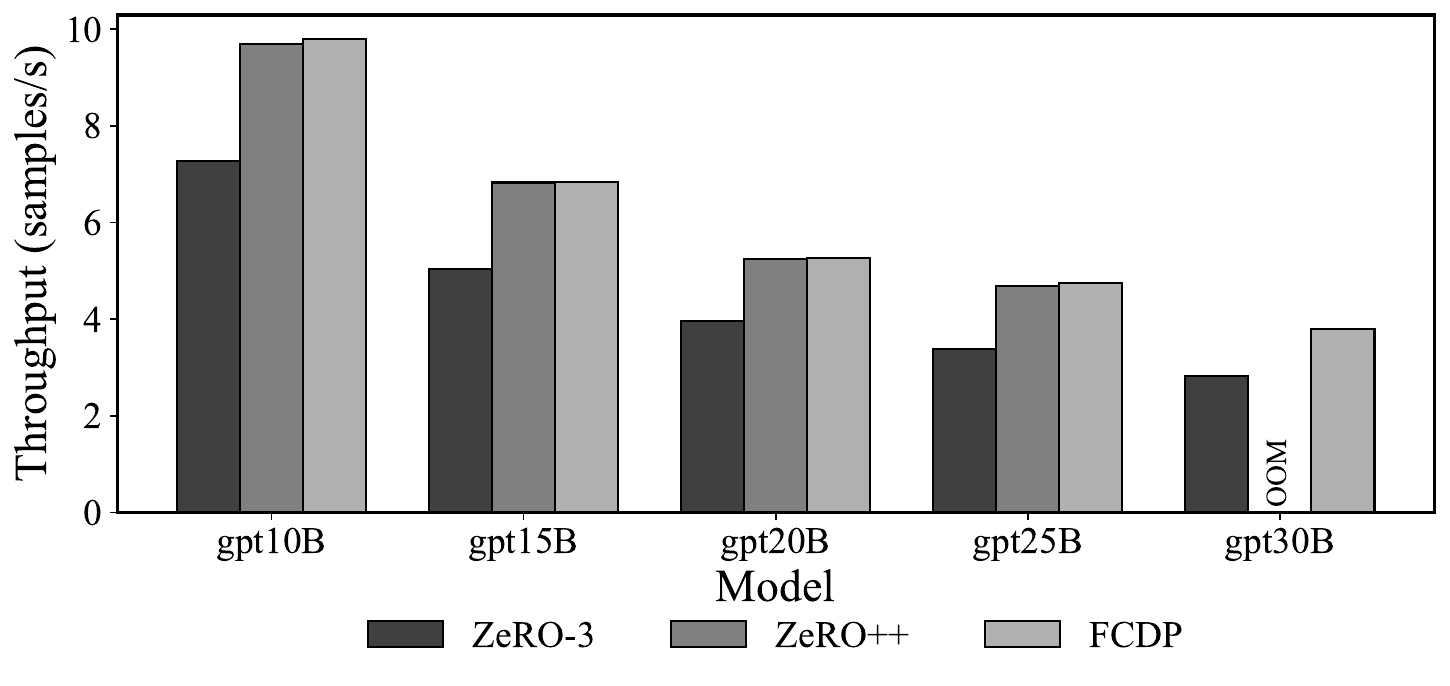}
}
\caption{Strong scaling performance across GPT models (10B--30B). ZeRO++ encounters out-of-memory failures on larger models (marked OOM).}
\label{graph:strong_scaling_result}
\end{figure*}

\subsection{Strong Scaling Performance}

We first evaluate strong scaling by fixing the microbatch size per GPU to 8 and varying the number of GPUs from 16 (two nodes) to 32 (four nodes).
Figure~\ref{graph:strong_scaling_result} reports training throughput in samples per second across five model sizes.
This experiment isolates communication overhead: with batch size held constant, throughput differences directly reflect each system's efficiency in reconstructing parameters across nodes.

\noindent\textbf{Result 1: \arch outperforms ZeRO-3 by up to 40.2\%.}
Across all configurations, \arch consistently delivers higher throughput than ZeRO-3.
This improvement stems from eliminating backward-pass inter-node \texttt{all-gather} by serving cached parameters from host memory.

\noindent\textbf{Result 2: \arch scales where ZeRO++ fails.}
The comparison with ZeRO++ reveals a fundamental trade-off.
When sufficient GPU memory is available (GPT-10B through GPT-20B on 32~GPUs), ZeRO++ achieves comparable throughput to \arch because both systems eliminate backward-pass inter-node \texttt{all-gather} through local caching.
However, ZeRO++ stores its cache in GPU memory, consuming capacity that could otherwise hold activations.
On GPT-20B with 16~GPUs (2 nodes) and GPT-30B with 32~GPUs (4 nodes), ZeRO++ triggers out-of-memory failures at a per-GPU batch size of 8, while \arch continues to operate without degradation.
This result demonstrates that host-memory caching preserves ZeRO++'s communication benefits without inheriting its memory constraints.

\begin{figure*}[t]
\centering
\subfloat[2nodes-16GPUs]
{
\includegraphics[
width=25em,
trim=6 0 0 0, 
clip
]{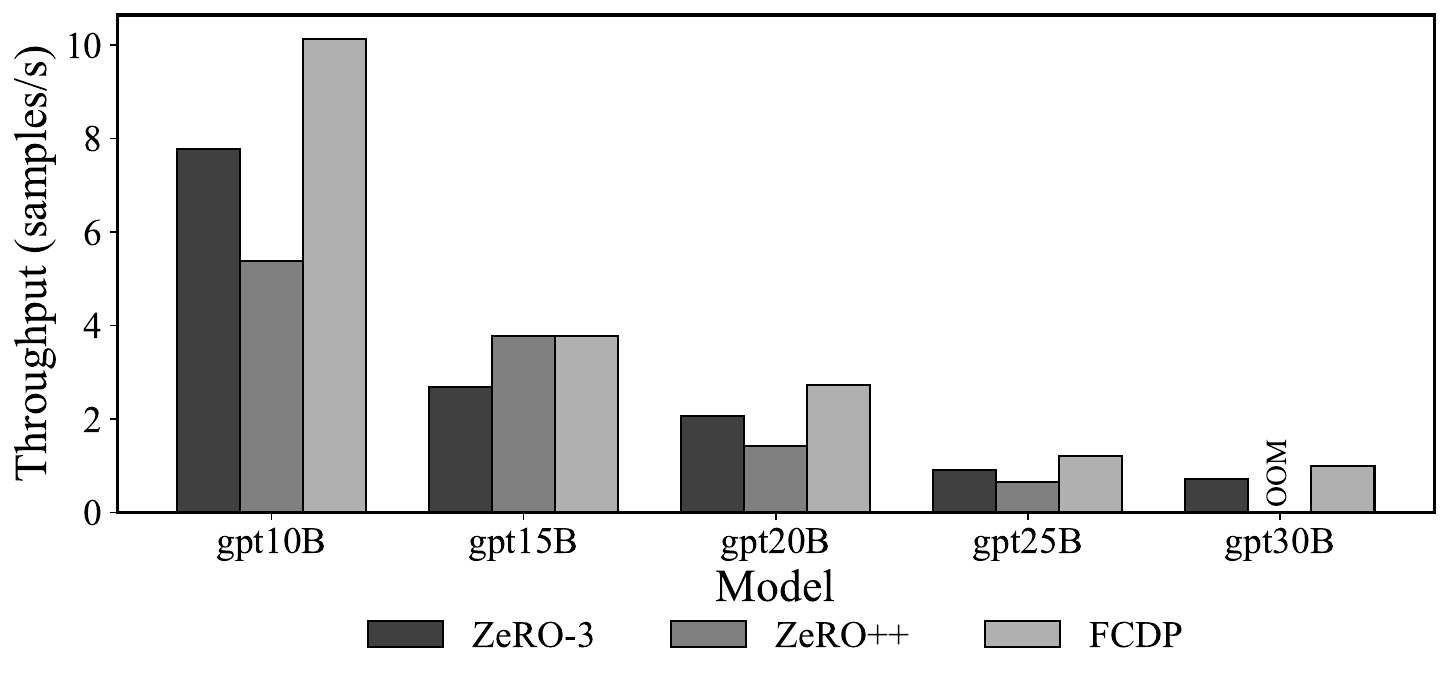}
}
\subfloat[4nodes-32GPUs]
{
\includegraphics[
width=25em,
trim=6 0 0 0, 
clip
]{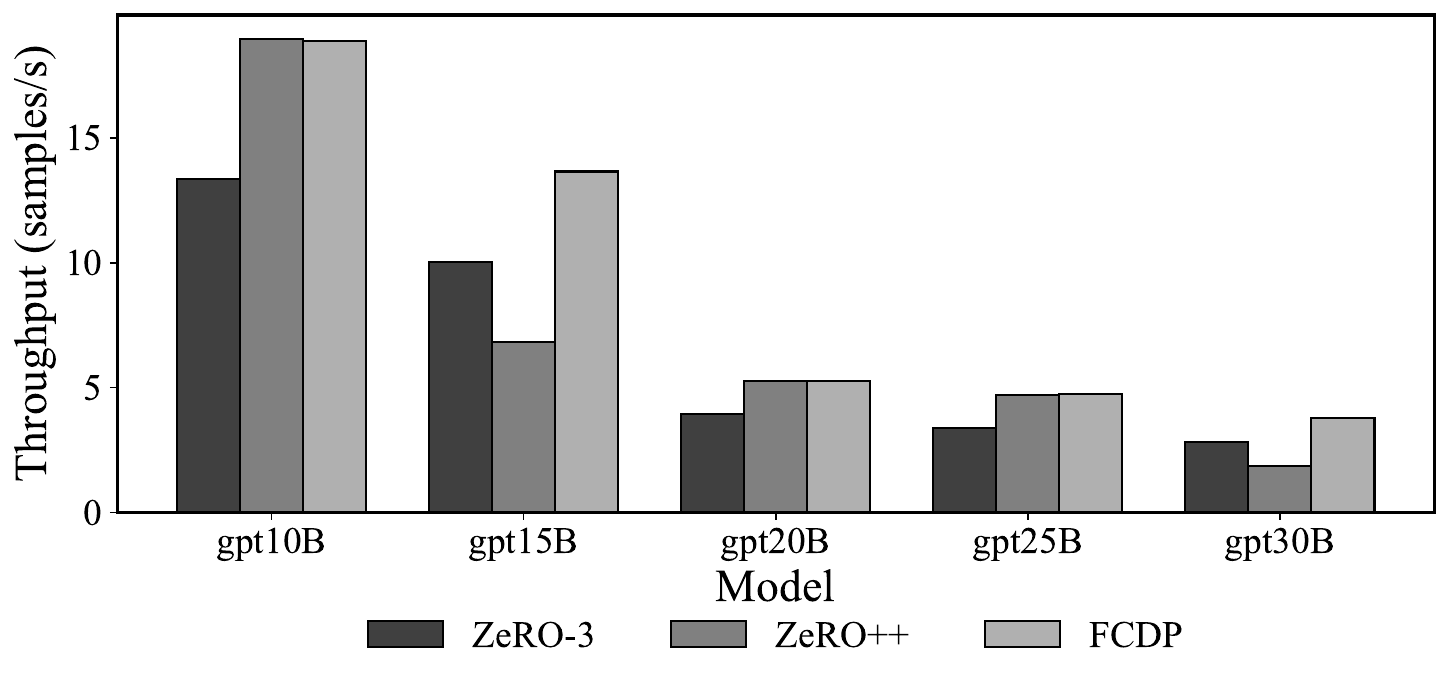}
}
\caption{Throughput under maximum batch size configurations. \arch maintains ZeRO-3--level memory efficiency while delivering communication-optimized performance.}
\label{graph:max_batch_result}
\end{figure*}

\subsection{Maximum Batch Size Performance}

We next evaluate throughput under memory pressure by maximizing the per-GPU batch size until GPU memory is exhausted.
For each configuration, we increase the per-GPU batch size by powers of two (1, 2, 4, 8, ...) until GPU memory is exhausted, then measure throughput at the largest successful batch size.
This approach maximizes hardware utilization by filling available GPU memory.
Figure~\ref{graph:max_batch_result} reports throughput under these memory-maximizing configurations; Tables~\ref{tab:max_batch_size_2nodes} and~\ref{tab:max_batch_size_4nodes} list the maximum batch sizes discovered for each system.

\noindent\textbf{Result 3: \arch matches ZeRO-3's batch capacity with up to 41.3\% higher throughput.}
\arch achieves the same maximum batch size as ZeRO-3 across all configurations because both systems have identical GPU memory footprints, since \arch's parameter cache resides in host memory.
At these matched batch sizes, \arch delivers up to 41.3\% higher throughput (GPT-10B, 32~GPUs) through reduced communication.

\noindent\textbf{Result 4: \arch achieves up to 2$\times$ higher throughput than ZeRO++.}
ZeRO++ supports smaller maximum batch sizes because its GPU-resident parameter cache competes with activations for limited GPU memory.
This memory pressure compounds into substantial throughput gaps.

\begin{table}[t]
\centering
\begin{tabular}{lccccc}
\hline
 & gpt10B & gpt15B & gpt20B & gpt25B & gpt30B \\
\hline
ZeRO-3  & 256 & 128 & 128 & 64 & 64 \\
ZeRO++ & 128 & 128 & 64  & 32 & OOM  \\
\arch   & 256 & 128 & 128 & 64 & 64 \\
\hline
\end{tabular}
\caption{Maximum batch size per model on 2-node configuration.}
\label{tab:max_batch_size_2nodes}
\end{table}

\begin{table}[t]
\centering
\begin{tabular}{lccccc}
\hline
 & gpt10B & gpt15B & gpt20B & gpt25B & gpt30B \\
\hline
ZeRO-3  & 512 & 512 & 256 & 256 & 256 \\
ZeRO++  & 512 & 256 & 256 & 256 & 128 \\
\arch    & 512 & 512 & 256 & 256 & 256 \\
\hline
\end{tabular}
\caption{Maximum batch size per model on 4-node configuration.}
\label{tab:max_batch_size_4nodes}
\end{table}

For smaller models (GPT-10B, GPT-15B), ZeRO++ operates successfully, though Tables~\ref{tab:max_batch_size_2nodes} and~\ref{tab:max_batch_size_4nodes} show it consistently supports equal or smaller batch sizes.
As model size grows, this memory tax becomes prohibitive.
On GPT-30B with 32~GPUs, \arch achieves \textbf{2$\times$ higher throughput} than ZeRO++.
On GPT-30B (2-node), ZeRO++ fails outright while \arch continues to operate at full capacity.

\noindent\textbf{Takeaway.}
These results demonstrate that \arch's host-memory caching decouples communication optimization from GPU memory capacity.
Host memory is abundant (512\,GB per node) and inexpensive compared to GPU memory (48\,GB per A40).
By exploiting this asymmetry, \arch captures ZeRO++'s communication benefits without inheriting its memory constraints.

\subsection{PEFT-Aware Communication Optimization}

We evaluate \arch on PEFT workloads using two nodes (16~GPUs).
We use LoRA adapters~\cite{hu2022lora} with rank $r=8$ applied to all attention projection layers (query, key, value, and output).
This configuration yields trainable parameters comprising $<$1\% of the total model, with the remaining $>$99\% frozen.
\archcomm exploits this structure by gathering frozen weights once at initialization and caching them in host memory indefinitely.
Subsequent iterations serve frozen parameters from local host memory via PCIe, while only the trainable adapter parameters require per-iteration inter-node \texttt{all-gather}.

\begin{figure}[t]
\centering
\includegraphics[width=1\linewidth]{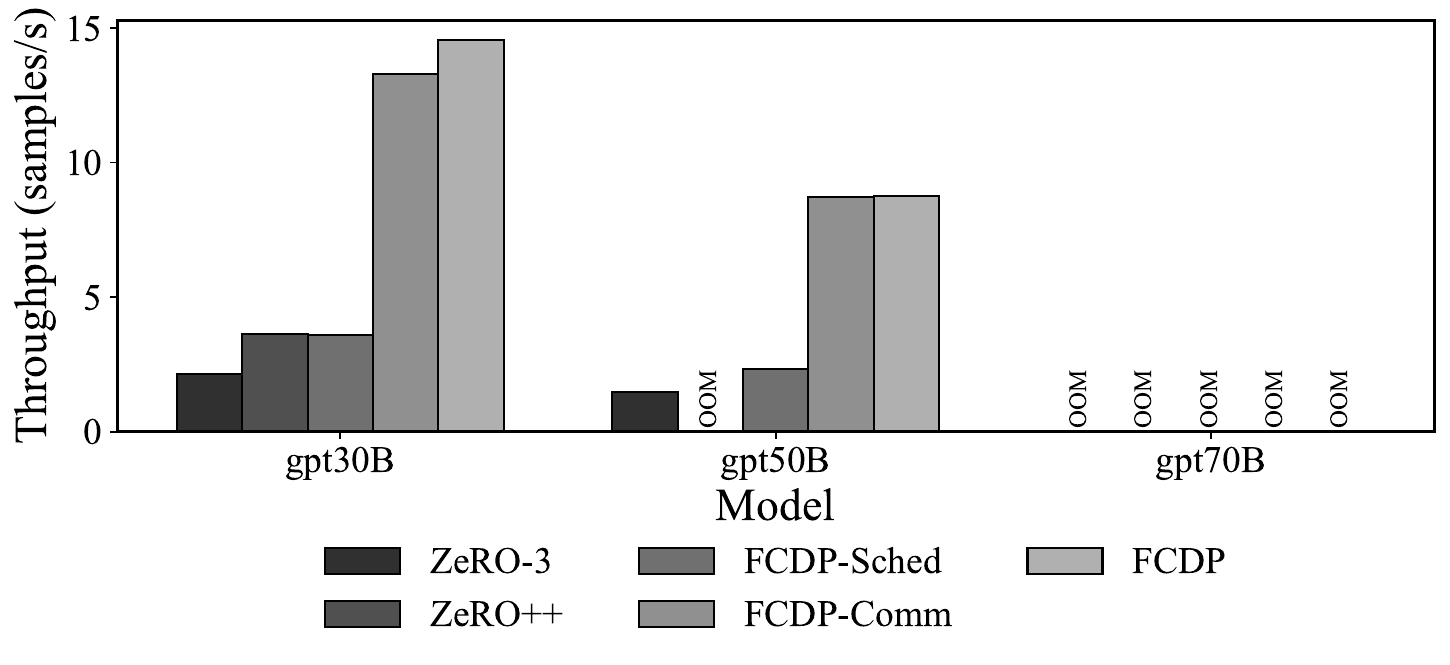}
\caption{PEFT strong scaling. \arch eliminates inter-node communication for frozen parameters, achieving higher throughput than ZeRO-3 and ZeRO++.}
\label{graph:strong_scaling_result_peft}
\end{figure}

\begin{figure}[t]
\centering
\includegraphics[width=1\linewidth]{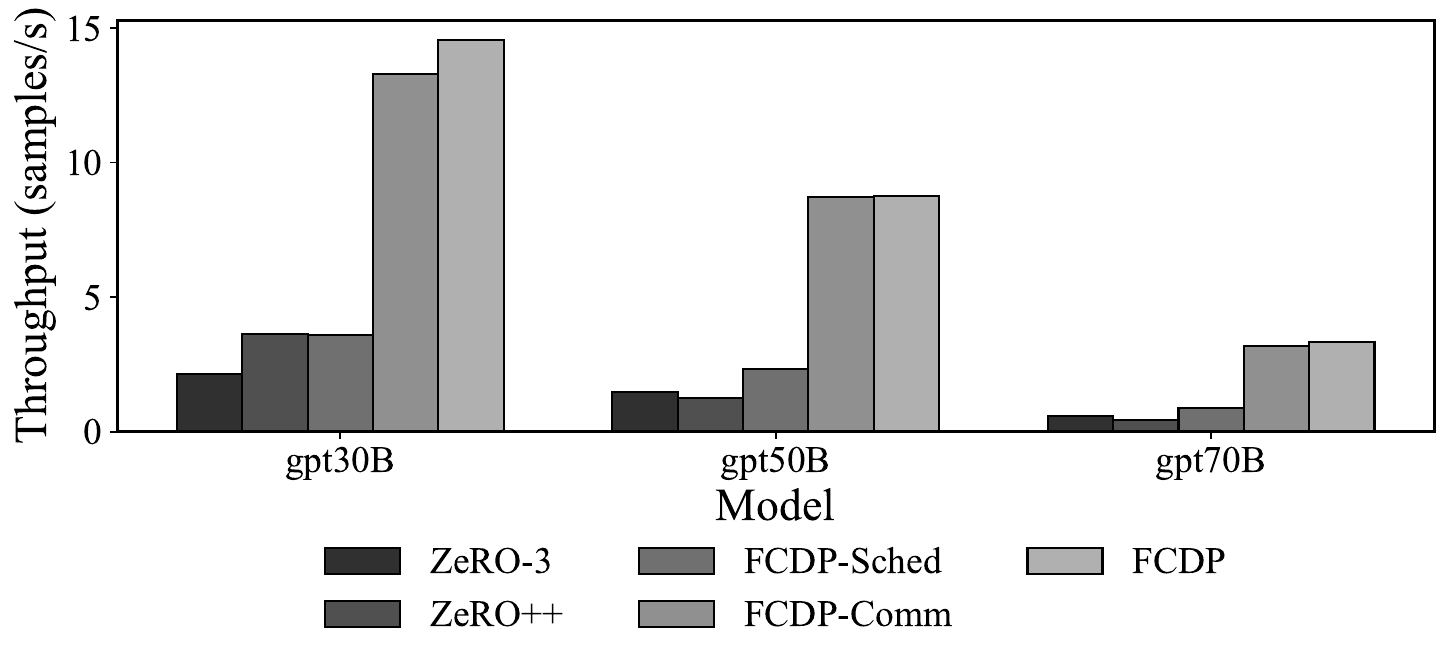}
\caption{PEFT throughput under maximum batch size configurations.}
\label{graph:max_batch_result_peft}
\end{figure}

\noindent\textbf{Result 5: \archcomm achieves up to 6.2$\times$ higher throughput than ZeRO-3 for PEFT.}
Figures~\ref{graph:strong_scaling_result_peft} and~\ref{graph:max_batch_result_peft} show PEFT training throughput under strong scaling and memory-maximizing configurations, respectively.
\archcomm consistently outperforms both ZeRO-3 and ZeRO++ across all model sizes.
The throughput gap is substantially larger than in full fine-tuning: whereas full fine-tuning reduces inter-node traffic by approximately 50\%, PEFT achieves a \textbf{99\%+ communication reduction} by also eliminating forward-pass transfers for frozen parameters.

\noindent\textbf{Result 6: \arch achieves up to 6.8$\times$ higher throughput than ZeRO-3 with GPU caching.}
As shown in Figures~\ref{graph:strong_scaling_result_peft} and~\ref{graph:max_batch_result_peft}, enabling \archmem further improves throughput by up to 6.8$\times$ compared to ZeRO-3.
The improvement is largest on GPT-30B, which has the most spare GPU memory available for caching.
Compared to ZeRO++, \arch achieves 4--7.4$\times$ higher throughput.

\begin{figure}[t]
\centering
\includegraphics[width=1\linewidth]{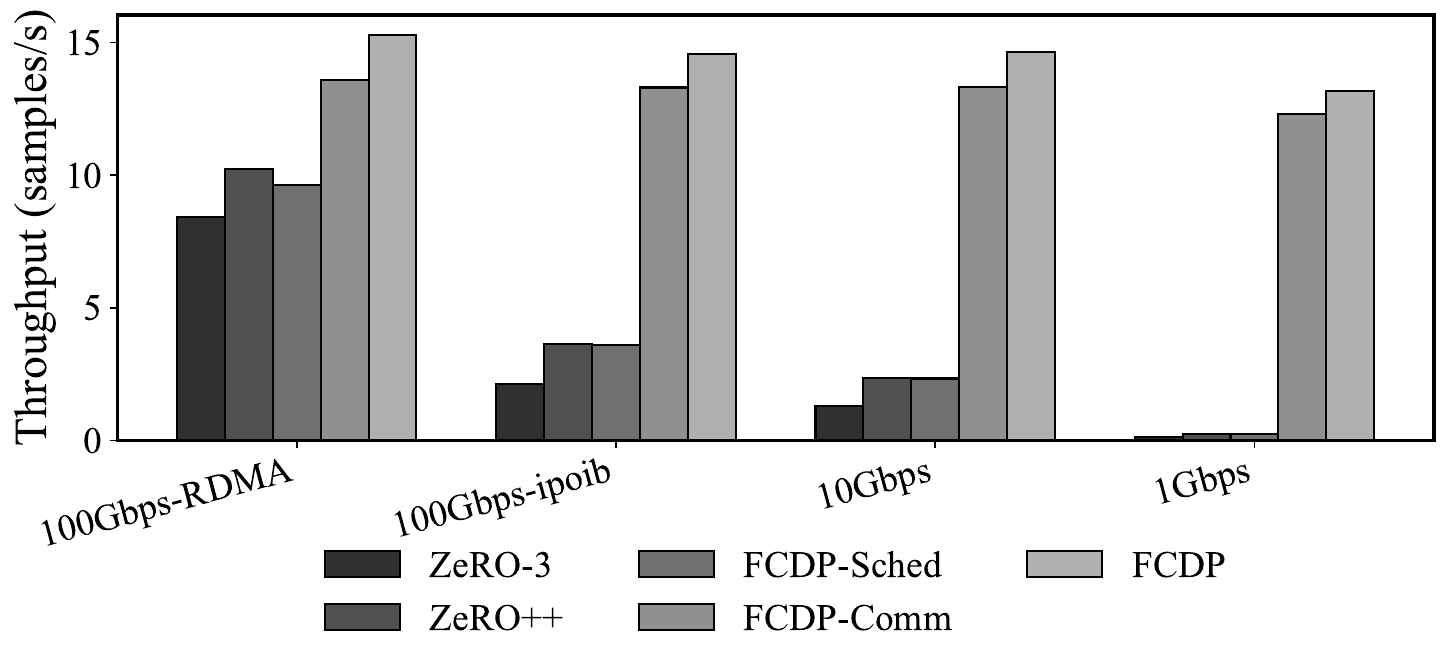}
\caption{Network bandwidth sensitivity for PEFT training. \arch achieves near-constant throughput by eliminating frozen parameter communication.}
\label{graph:peft_network}
\end{figure}

\noindent\textbf{Result 7: \arch is insensitive to network bandwidth for PEFT.}
Figure~\ref{graph:peft_network} examines sensitivity to network bandwidth by varying inter-node connectivity from 100\,Gbps InfiniBand to 1\,Gbps Ethernet.
Because 99\% of parameters require no inter-node communication, \archcomm maintains 90.5\% and \arch maintains 86.3\% of peak throughput even on 1\,Gbps Ethernet.
In contrast, ZeRO-3 throughput degrades by 98.4\% and ZeRO++ by 97.4\% as bandwidth decreases from 100\,Gbps to 1\,Gbps.
As a result, \arch achieves up to 100$\times$ higher throughput than ZeRO-3 and 51$\times$ higher than ZeRO++.
This result confirms that \archcomm effectively decouples PEFT training performance from inter-node network capacity.

\begin{figure*}[t]
\centering
\includegraphics[width=1\linewidth]{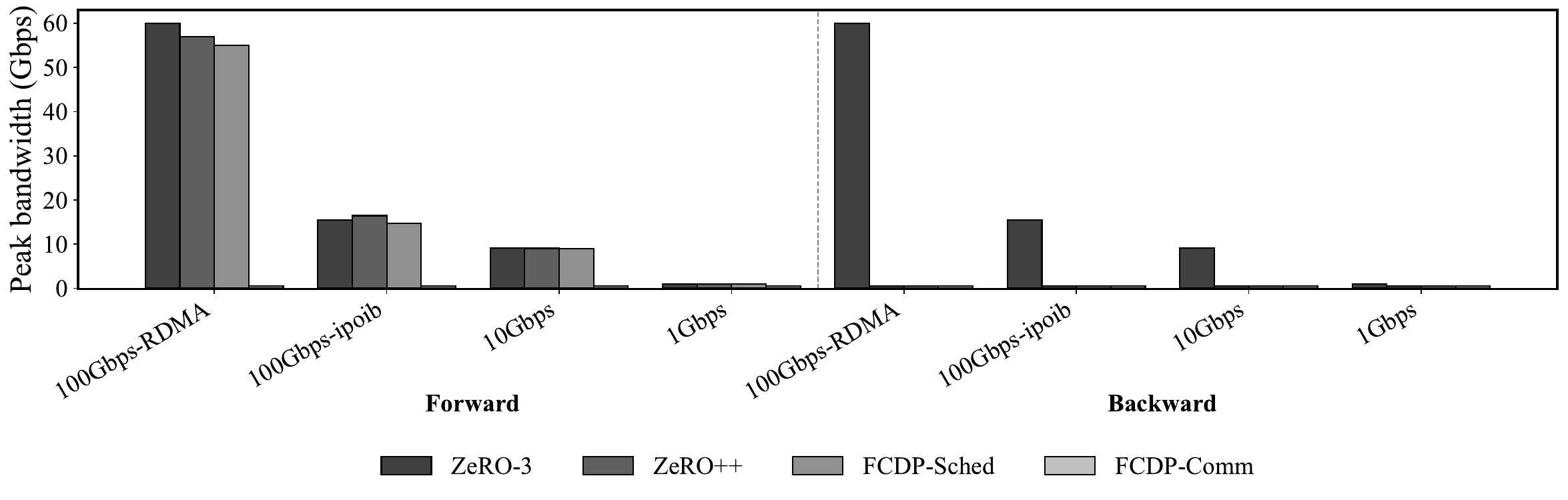}
\caption{Peak inter-node network bandwidth during forward and backward passes. \arch's communication reduction results in lower sustained network load compared to ZeRO-3.}
\label{graph:realtime_bw}
\end{figure*}

\subsection{Communication Volume Analysis}

To understand where throughput improvements originate, we measure inter-node communication volume during PEFT training.
Table~\ref{tab:inter_node_comm_volume} breaks down bytes transferred during forward and backward passes per iteration.

\begin{table}[t]
\centering
\begin{tabular}{lccc}
\hline
 & Forward & Backward & Total \\
\hline
ZeRO-3        & 110.25 & 103.48 & 213.73 \\
ZeRO++       & 108.18 &  0.1 &  108.28 \\
\archflow   & 108.18 &  0.1 &  108.28 \\
\archcomm    & 0.06 &  0.1 &   0.16 \\
\hline
\end{tabular}
\caption{Inter-node communication volume (GB) during forward and backward passes per iteration.}
\label{tab:inter_node_comm_volume}
\end{table}

\noindent\textbf{Result 8: \archflow reduces communication by 49.3\%, \archcomm by 99.9\%.}
ZeRO-3 transfers 213.73\,GB per iteration, split roughly evenly between forward and backward passes.
\archflow reduces this to 108.28\,GB (\textbf{49.3\% reduction}) by eliminating backward-pass inter-node \texttt{all-gather}. Parameters cached in host memory are reconstructed via fast intra-node transfers instead.
\archcomm further reduces communication to 0.16\,GB (\textbf{99.9\% reduction}) by caching frozen parameters and transferring only trainable adapter weights.

Figure~\ref{graph:realtime_bw} visualizes peak inter-node network bandwidth during forward and backward passes.
The trace confirms that \arch's communication reduction translates to lower sustained network load, freeing network capacity for other cluster workloads.

\section{Discussion}
\label{sec:discussion}

This section provides a theoretical analysis of the memory and communication trade-offs that motivated \arch's design.
We derive the memory footprint and communication volume for each system, then show how \arch navigates the design space to achieve both efficiency and scalability.

\subsection{Memory Allocation Analysis}

We begin by establishing notation.
Let $W$ denote the total number of model parameters, $G$ the total number of GPUs across the cluster, $N$ the number of nodes, and $g = G/N$ the number of GPUs per node.
For concreteness, consider a 30B-parameter model ($W = 30 \times 10^9$) distributed across 4 nodes with 8 GPUs each ($G = 32$, $N = 4$, $g = 8$).

\noindent\textbf{ZeRO-3.}
ZeRO-3 partitions parameters evenly across all $G$ GPUs, so each GPU stores a shard of size $W/G$.
In our running example, this yields $30\text{B}/32 \approx 0.94\text{B}$ parameters per GPU, or approximately 1.9\,GB in FP16.
During forward and backward passes, each layer's full parameters must be reconstructed before computation.
ZeRO-3 performs this reconstruction via \texttt{all-gather}, uses the parameters for one computation step, then immediately discards them to reclaim memory.
This strategy achieves the minimum possible GPU memory footprint (only $W/G$ parameters reside on each GPU at any time), but it requires two \texttt{all-gather} operations per layer per iteration, one for forward and one for backward.

\noindent\textbf{MiCS.}
MiCS partitions GPUs into subgroups of size $S$ and shards parameters only within each subgroup.
When $S = g$ (one subgroup per node), each node collectively holds a complete copy of all parameters, enabling parameter reconstruction via fast intra-node \texttt{all-gather} rather than slow inter-node communication.
However, this comes at a memory cost: each GPU now stores $W/S = W/g$ parameters instead of $W/G$.
In our example, GPU memory increases from 0.94B to $30\text{B}/8 = 3.75\text{B}$ parameters (7.5\,GB in FP16), a $4\times$ increase corresponding to the number of nodes $N = G/g$.
This additional memory pressure reduces the batch size that fits in GPU memory, limiting throughput.

\noindent\textbf{ZeRO++.}
ZeRO++ takes a hybrid approach: it maintains ZeRO-3's global sharding (each GPU stores $W/G$) but additionally caches a secondary full-parameter replica at the node level.
This cache is shared across the $g$ GPUs within a node, with each GPU holding $W/g$ cached parameters.
The effective GPU memory footprint becomes $W/G + W/g$.
Since $W/G \ll W/g$ for typical configurations ($G \gg g$), the footprint is dominated by the cache term: approximately $W/g$, the same as MiCS.
In our example, each GPU requires $\approx$7.5\,GB for the cache alone, leaving less memory for activations and reducing maximum batch size.

\noindent\textbf{\arch.}
\arch stores the same cached replica in host memory instead of GPU memory.
GPU memory remains exactly $W/G$, identical to ZeRO-3, while host memory holds $W$ per node (approximately $2W$ bytes in FP16, or 60\,GB for a 30B model).
This trade-off is highly favorable on commodity servers, which typically provide 512\,GB--1\,TB of host memory versus 48--80\,GB per GPU.
By offloading the cache to the abundant host memory tier, \arch preserves ZeRO-3's GPU memory efficiency while enabling the communication benefits that MiCS and ZeRO++ achieve only at the cost of increased GPU memory.

\subsection{Communication Volume Analysis}

We now derive the inter-node communication volume per training iteration, showing how each system's design choices affect network traffic.

\noindent\textbf{Forward-pass all-gather.}
Before computing the forward pass for layer $l$, the system must reconstruct the layer's full parameters $W_l$ from shards distributed across GPUs.
In ZeRO-3, this reconstruction requires an \texttt{all-gather} operation across all $G$ GPUs.
For inter-node traffic analysis, we note that each GPU receives $(G-g)/G$ of the parameters from remote nodes (the remaining $g/G$ comes from local GPUs via fast intra-node transfers).
Simplifying, inter-node traffic per layer is approximately $(N-1)/N \cdot W_l$ bytes.
Summing across all layers, the forward-pass inter-node traffic is:
\[
\text{Fwd-AG} = \frac{N-1}{N} \cdot W \approx W \quad \text{for } N \geq 2
\]
This cost is identical for ZeRO-3, MiCS, ZeRO++, and \arch, since all systems must reconstruct parameters before forward computation.

\noindent\textbf{Backward-pass all-gather.}
During the backward pass, gradients must be computed with respect to each layer's parameters, requiring the full parameter tensor to be available again.
ZeRO-3 performs another \texttt{all-gather} to reconstruct parameters, incurring the same inter-node traffic as the forward pass:
\[
\text{Bwd-AG}_{\text{ZeRO-3}} = \frac{N-1}{N} \cdot W \approx W
\]
ZeRO++ and \arch eliminate this cost entirely by caching parameters after the forward pass.
ZeRO++ serves the backward pass from its GPU-resident cache; \arch serves it from host memory via PCIe transfers.
In both cases, the backward-pass reconstruction uses only local resources:
\[
\text{Bwd-AG}_{\text{ZeRO++}} = \text{Bwd-AG}_{\arch} = 0
\]

\noindent\textbf{Gradient reduce-scatter.}
After the backward pass, gradients must be aggregated across all GPUs and redistributed so each GPU holds the gradient shard corresponding to its parameter shard.
This \texttt{reduce-scatter} operation incurs $(N-1)/N \cdot W \approx W$ bytes of inter-node traffic.
This cost is unavoidable in data-parallel training and identical across all systems.

\noindent\textbf{Total per-iteration communication.}
Combining the three phases, we obtain the total inter-node communication per iteration:
\begin{align*}
\text{ZeRO-3:} \quad & W + W + W = 3W \\
\text{ZeRO++/\arch:} \quad & W + 0 + W = 2W
\end{align*}
By caching parameters after the forward pass, ZeRO++ and \arch reduce inter-node traffic by 33\% compared to ZeRO-3.
The key difference is where the cache resides: ZeRO++ uses GPU memory (increasing memory pressure), while \arch uses host memory (preserving GPU capacity).

\subsection{PEFT Communication Analysis}

PEFT workloads present an additional optimization opportunity that \arch exploits.
In PEFT, only a small subset of parameters $W_t$ (typically $<$1\% of the model) are trainable; the remaining $W_f = W - W_t$ are frozen throughout training.

\noindent\textbf{Frozen parameter caching.}
Since frozen parameters never change, they need only be gathered once at initialization and can then be cached permanently.
\arch gathers $W_f$ at startup and stores it in host memory, where it remains for the duration of training.
Subsequent iterations serve $W_f$ directly from local host memory, incurring zero inter-node traffic for these parameters.
Only the trainable parameters $W_t$ require per-iteration \texttt{all-gather}.

\noindent\textbf{PEFT communication volume.}
For ZeRO-3, which is unaware of the frozen/trainable distinction, communication remains proportional to total model size:
\[
\text{ZeRO-3 (PEFT):} \quad W + W + W_t = 2W + W_t \approx 2W
\]
For \archcomm, forward and backward passes transfer only trainable parameters:
\[
\text{\archcomm (PEFT):} \quad W_t + 0 + W_t = 2W_t
\]
For LoRA with $W_t = 0.01W$, this represents a $\approx$100$\times$ reduction in inter-node traffic compared to ZeRO-3.
This optimization is particularly impactful because PEFT has become the dominant paradigm for adapting large language models to downstream tasks.

\subsection{Design Space Trade-offs}

The preceding analysis reveals three orthogonal design dimensions that differentiate distributed training systems.

\noindent\textbf{Cache location: GPU vs.\ host memory.}
ZeRO++ caches parameters in GPU memory, providing the lowest-latency access during backward-pass reconstruction.
However, this cache consumes memory that could otherwise hold activations, reducing maximum batch size.
\arch caches in host memory, accepting higher PCIe latency in exchange for preserving GPU memory capacity.
Because PCIe transfers overlap with layer computation, this latency is largely hidden.
On commodity clusters where host memory is 10$\times$ larger than GPU memory, \arch's trade-off enables training larger models at larger batch sizes.

\noindent\textbf{Sharding granularity: global vs.\ subgroup.}
MiCS shards within subgroups to confine \texttt{all-gather} operations to fast intra-node communication.
However, subgroup sharding increases per-GPU memory by factor $N$, the same cost as explicit caching.
\arch maintains global sharding (identical to ZeRO-3) for minimum memory footprint, achieving MiCS's communication benefits through host-memory caching rather than subgroup replication.

\noindent\textbf{Workload awareness: uniform vs.\ PEFT-aware.}
All prior systems (ZeRO-3, MiCS, ZeRO++) treat parameters uniformly, transferring all weights every iteration regardless of whether they are trainable.
\arch distinguishes frozen from trainable parameters at runtime, enabling order-of-magnitude communication reduction for PEFT workloads.
This workload-aware optimization is increasingly important as PEFT becomes the standard approach for LLM deployment.

\section{Related Work}
\label{sec:related}

\noindent\textbf{Sharded Data Parallelism.}
ZeRO~\cite{rajbhandari2020zero} partitions optimizer states (ZeRO-1), gradients (ZeRO-2), or all model states (ZeRO-3) across GPUs, reconstructing full parameters via \texttt{all-gather} on demand.
ZeRO-3 achieves the lowest memory footprint but performs two inter-node \texttt{all-gather} operations per layer per iteration (one for the forward pass and one for the backward pass), yielding $3W$ bytes of inter-node traffic per step.
PyTorch Fully Sharded Data Parallel (FSDP)~\cite{zhao2023fsdp} provides equivalent sharding with tighter autograd integration, enabling automatic resharding at module boundaries and native support for mixed-precision training.
PyTorch DDP~\cite{li2020pytorch} overlaps bucketed \texttt{all-reduce} gradient synchronization with backward computation, a pattern that FSDP extends to full parameter sharding.
However, both ZeRO-3 and FSDP treat host memory purely as overflow storage and reconstruct parameters from remote shards in both forward and backward passes.
Unlike these systems, \arch caches parameters in host memory after the forward-pass \texttt{all-gather}, eliminating backward-pass inter-node traffic entirely and reducing per-iteration communication from $3W$ to $2W$.

\noindent\textbf{Model Parallelism and 3D Parallelism.}
Tensor parallelism~(TP)~\cite{shoeybi2019megatron, shazeer2018mesh} partitions individual operators across GPUs within a node, requiring \texttt{all-reduce} after each partitioned layer.
Pipeline parallelism~(PP) partitions layers across pipeline stages: GPipe~\cite{huang2019gpipe} accumulates micro-batches to fill pipeline bubbles, while PipeDream~\cite{narayanan2019pipedream} overlaps forward and backward passes across stages to improve hardware utilization.
Megatron-3D~\cite{narayanan2021megatron3d} combines TP, PP, and data parallelism (DP) into a unified 3D parallelism framework, demonstrating efficient scaling to thousands of GPUs.
However, TP requires high-bandwidth intra-node interconnects (NVLink or NVSwitch) and PP demands model-specific stage partitioning~\cite{athlur2022varuna}, making both approaches hardware- and model-dependent.
Unlike TP and PP, \arch operates entirely within the DP dimension, requiring no model-specific partitioning.
\arch is orthogonal to TP and PP and can serve as the DP component within a 3D parallelism configuration.

\noindent\textbf{Communication Optimization.}
MiCS~\cite{zhang2022mics} confines \texttt{all-gather} within GPU subgroups, trading memory for reduced communication scope.
ZeRO++~\cite{wang2023zero++} caches node-local parameter replicas in GPU memory, eliminating backward-pass inter-node traffic but increasing GPU memory pressure by $W/g$ per device.
Horovod~\cite{sergeev2018horovod} popularized bandwidth-optimal ring \texttt{all-reduce} for gradient synchronization, and BytePS~\cite{jiang2020byteps} extends this to heterogeneous CPU--GPU clusters by leveraging spare CPU and bandwidth resources as parameter servers.
BAGUA~\cite{gan2022bagua} provides a modular framework supporting asynchronous communication, decentralized gradient exchange, and communication--computation overlap.
TACCL~\cite{shah2023taccl} synthesizes topology-aware collective algorithms by exploiting the physical network graph, and NCCL~\cite{nccl2023} provides highly optimized GPU-native collective primitives.
TCCL~\cite{kim2024tccl} addresses NCCL's suboptimal performance on PCIe GPU clusters by discovering optimized communication paths.
However, these systems optimize collective \emph{algorithms} (how data is routed) without reducing the communication \emph{volume} itself.
Unlike algorithmic optimizations, \arch reduces the total bytes transferred per iteration from $3W$ to $2W$, a complementary improvement that compounds with any underlying collective implementation.

\noindent\textbf{Heterogeneous Memory Management.}
ZeRO-Offload~\cite{ren2021zerooffload} offloads optimizer states and gradients to host memory, enabling single-GPU training of billion-parameter models.
ZeRO-Infinity~\cite{rajbhandari2021zeroinfinity} extends offloading to NVMe storage, supporting trillion-parameter models by exploiting the full memory hierarchy.
Capuchin~\cite{peng2020capuchin} performs tensor-level swap scheduling between GPU and host memory, using access-pattern prediction to prefetch evicted tensors before reuse.
PatrickStar~\cite{fang2022patrickstar} manages model data in chunks that migrate dynamically between GPU and host memory based on runtime memory statistics collected during a warm-up iteration.
Harmony~\cite{li2022harmony} partitions DNN layers across commodity server resources, co-optimizing host memory placement and PCIe transfer scheduling.
All of these systems treat host memory as a \emph{capacity extension}, a slower tier that absorbs data when GPU memory is exhausted.
Unlike capacity-driven offloading, \arch uses host memory as a \emph{communication cache}: parameters are staged in host DRAM not because GPU memory overflows, but because doing so eliminates an inter-node \texttt{all-gather}, converting a network-bound operation into a local PCIe transfer.

\noindent\textbf{PEFT-Aware Distributed Training.}
LoRA~\cite{hu2022lora} freezes pre-trained weights and injects trainable low-rank matrices into selected Transformer layers, reducing trainable parameters to $<$1\% of the model.
Adapters~\cite{houlsby2019adapters} insert small bottleneck modules within each transformer layer, and Prefix-Tuning~\cite{li2021prefix} prepends learnable continuous vectors to each layer's input sequence.
These methods have become the dominant paradigm for LLM fine-tuning, yet existing distributed frameworks (ZeRO-3 and FSDP) remain oblivious to the frozen/trainable distinction and transfer all parameters every iteration.
Unlike prior frameworks, \arch identifies frozen parameters at runtime and caches them permanently in host memory, performing inter-node \texttt{all-gather} only for the trainable subset $W_t$.
For LoRA with $W_t \approx 0.01W$, this reduces per-iteration inter-node traffic from $2W$ to $2W_t$, a $\approx$100$\times$ reduction.

\noindent\textbf{Commodity Cluster Training.}
Fire-Flyer~\cite{an2024fireflyer} demonstrates cost-effective HPC through PCIe-based GPU interconnect and software-hardware co-design, achieving comparable performance to DGX systems at substantially lower cost.
Analysis of production cluster traces from Microsoft Philly~\cite{jeon2019analysis} and Alibaba PAI~\cite{weng2022mlaas} confirms that commodity GPU clusters are the dominant deployment environment for deep learning workloads.
However, Fire-Flyer's software stack (HFReduce, 3FS, HAI-Platform) is tightly coupled to its specific cluster configuration, limiting portability to other installations.
Unlike hardware co-design approaches, \arch is a software-only solution that deploys on any commodity cluster without hardware modifications, leveraging only the host memory and PCIe bandwidth already present in standard server configurations.

\section{Conclusions}


%
Distributed training of large language models via fully sharded data parallelism (ZeRO-3) suffers from severe communication bottlenecks on commodity clusters where inter-node bandwidth is limited.
Existing approaches either cache parameters in GPU memory (sacrificing memory capacity) or offload to host memory for capacity extension (degrading throughput).
This paper presented \arch, a distributed training framework that uses host memory as a fast caching layer, not an overflow tier, to reduce inter-node communication while preserving ZeRO-3's minimal GPU memory footprint.

\arch introduces three techniques.
First, \archflow exploits the redundancy that ZeRO-3 gathers identical parameters twice per layer: it caches forward-pass parameters in host memory and reconstructs them via fast intra-node \texttt{all-gather} during the backward pass, reducing inter-node traffic by 50\%.
Second, \archmem monitors GPU memory pressure and adaptively places parameters on-device when headroom exists or in host memory under pressure, guaranteeing worst-case memory usage identical to ZeRO-3 while opportunistically eliminating PCIe transfers.
Third, \archcomm classifies parameters into frozen and trainable at initialization; frozen weights are gathered once and cached indefinitely, restricting inter-node \texttt{all-gather} to trainable adapters only. This reduces communication by over 99\% for LoRA fine-tuning.
On commodity clusters where communication previously dominated iteration time, \arch achieves up to 41.3\% higher throughput than ZeRO-3 for full fine-tuning and 100$\times$ speedup for PEFT workloads, while matching ZeRO-3's maximum batch size.
By demonstrating that host memory can outperform inter-node communication on bandwidth-limited clusters, \arch enables efficient large model training on commodity hardware without sacrificing memory efficiency.

\section*{Acknowledgment}
This work was supported by Institute of Information \& communications Technology Planning \& Evaluation (IITP) grant funded by the Korea government (MSIT) (No.2022-0-00498; No.RS-2025-02263167). The corresponding author is Ki-Dong Kang (kd\_kang@etri.re.kr).

\bibliographystyle{IEEEtranS}
\bibliography{references}

\end{document}